\documentclass{natureprintstyle}
\usepackage[T1]{fontenc}
\usepackage{times}

\usepackage{graphicx}
\usepackage{amsmath, amssymb}

\title{Arsenene: Two-dimensional buckled and puckered honeycomb arsenic systems}

\author{C. Kamal$^1$  and Motohiko Ezawa$^2$}

\begin{document}

\maketitle

\begin{affiliations}
 \item Indus Synchrotrons Utilization Division, Raja Ramanna Centre for Advanced
Technology, Indore 452013, India,
 \item Department of Applied Physics, University of Tokyo, Hongo 7-3-1, 113-8656, Japan
\end{affiliations}

\begin{abstract}
Recently phosphorene, monolayer honeycomb structure of black phosphorus, was experimentally manufactured and attracts rapidly growing interests. Here we investigate stability and electronic properties of honeycomb structure of arsenic system based on first principle calculations.  Two types of honeycomb structures, buckled and puckered, are found to be stable. We call them arsenene as in the case of phosphorene. We find that both the buckled and puckered arsenene possess indirect gaps. We show that the band gap of the puckered and buckled arsenene can be tuned by applying strain. The gap closing occurs at 6\% strain for puckered arsenene, where the bond angles between the nearest neighbour become nearly equal. An indirect-to-direct gap transition occurs by applying strain. Especially, $1\%$ strain is enough to transform the puckered arsenene into a direct-gap semiconductor. Our results will pave a way for applications to light-emitting diodes and solar cells.
\end{abstract}

Graphene, a planar honeycomb monolayer of carbon atoms, is one of the most fascinating materials\cite{NetoRev,KatsText}. It has high mobility, heat conductance and mechanical strength. However, it lacks an intrinsic band gap, which makes electronic applications of graphene difficult. The finding of graphene excites material search of other monolayer honeycomb systems with intrinsic gaps.  Recently, honeycomb structures of the carbon group attract much attention, which are silicene, germanene and stanene\cite{LiuPRB}. The geometric structures of these systems are buckled due to the hybridization of sp$^2$ and sp$^3$ orbitals. Accordingly we can control the band gap by applying perpendicular electric field\cite{Falko,EzawaNJP,Kamal}. These are topological insulators owing to spin-orbit interactions\cite{LiuPRL}. Although silicene and germanene have already been manufactured on substrates\cite{GLay,Kawai,Takamura}, their free-standing samples are not yet available, which makes experiments difficult to reveal their exciting properties. Phosphorene, a monolayer of black phosphorus, was recently manufactured by exfoliating black phosphorus\cite{Li,Liu,Xia,Gomez,Koenig,Bus}. It has already been shown that it acts as a field-effect transistor\cite{Li}. The experimental success evokes recent flourish of studies of phosphorene\cite{Reich, Fei,Peng,Rodin,Wei,Fei2,Qiao}. The structure is puckered, which is different from the planar graphene and the buckled silicene. Furthermore, the buckled phosphorene named "blue phosphorene" is shown to be stable by first principle calculations\cite{Blue}.

In this paper, motivated by recent studies on phosphorene, we have investigated stability and electronic properties of arsenene, which is a honeycomb monolayer of arsenic, by employing density functional theory (DFT) based electronic structure calculations. First we show two types of honeycomb structures, namely buckled and puckered, are stable by investigating phonon spectrum and cohesive energy. Our calculations show that the buckled arsenene is slightly more stable than the puckered arsenene. Though both these two systems possess indirect band gaps, it is possible to make a transition from an indirect to direct band gap by applying strain or external electric field. Puckered arsenene is transformed into a direct-gap semiconductor by applying only $1\%$ strain, and the gap nearly closes at $6\%$ strain.
The band gap of the buckled arsenene can be tuned by  electric field, while the band gap change is negligible for the puckered arsenene.

\bigskip
\noindent{\large\bf RESULTS}

\section*{Stability of Arsenene.}
\quad

\begin{table*}[t]
\begin{center}
\caption{The results for optimized geometries of arsenene obtained by DFT with PBE exchange-correlations functional.}
 \begin{tabular}{lccccccc}
 \hline
\hline
Structure	&	Space&  Cohesive&\multicolumn{2}{c}{Lattice constants (\AA{})} & Bond & Bond \\  \cline{4-5}
		& group 	&energy (eV/atom)&$\textbf{a}$	&$\textbf{b}$ or $\textbf{c}$&  length (\AA{})  & angle  ($^\circ$)  \\
\hline
Puckered	&Pmna	&-2.952	&3.677	&4.765 ($\textbf{b}$)	&2.501, 2.485	&100.80, 94.64\\
Planar	&P6/mmm	&-2.391	&4.366	&-				&2.521		&120.00 \\
Buckled	&P3m1	&-2.989	&3.607	&-				&2.503		&92.22 \\
Bulk		& R-3m	&-2.986	&3.820  & 10.752 ($\textbf{c}$)	&2.556 &96.72 \\
		&		&		&(3.7598)\cite{As-expt}&(10.5475)\cite{As-expt}&&	\\
\hline
\hline
\end{tabular}
\end{center}
\end{table*}

Graphene forms a planar honeycomb structure since it exhibits purely sp$^2$ hybridization. On the other hand, other elemental honeycomb systems so far found are not planar but form either buckled or puckered structure. For example, a honeycomb structure of group IV element such as silicene, germanene and stanene form a buckled structure. Additionally, phosphorene made of phosphorus belonging to group V is known experimentally to have a puckered structure. It has theoretically been shown that there is also a buckled structure of phosphorene named blue phosphorene\cite{Blue}. These observations make it important to study if there is a  stable honeycomb structure made of arsenic -  another group V element.
For this purpose, we choose three different possible honeycomb geometric structures, namely (i) puckered, (ii) planar and (iii) buckled for arsenene. We show the optimized geometric structures for these three cases in Figure 1(a),(b),(c). The results of optimized geometric structures are also summarized in Table 1. 
The puckered angle of arsenene is 100.80$^\circ$, which is slightly smaller than that of phosphorene 103.69$^\circ$\cite{Gomez}. In the case of the buckled structure, the buckling height and angle are found to be 1.388 \AA{} and 92.22$^\circ$, respectively.

In order to study the stability of arsenene, we have carried out the cohesive energy as well as phonon dispersion calculations for the above mentioned three possible structures. 
The cohesive energy of -2.952, -2.391 and -2.989 eV/atom for the puckered, planar and buckled arsenene, respectively. Among the three two-dimensional structures, the buckled arsenene is the minimum energy configuration. 
However, the cohesive energy difference between the buckled and puckered systems is very small and it is comparable with the thermal energy at the room temperature. On the other hand, the cohesive energy of planar structure is about 400 meV less as compared to those of the other two structures. 

Furthermore, we have performed the phonon dispersion calculations for these three systems. The results of phonon dispersion along the high symmetric points in the Brillouin zone (See Figure 1(d), (e), (f))  for these three systems are given in Figure 1(g), (h), (i). From the phonon spectrum, it is possible to compare the stability and structural rigidity of these systems.
Puckered arsenene is globally stable since the global energy minimum exists, and the phonon dispersion is completely positive and linear around the $\Gamma$ point . In the case of buckled arsenene, all the modes contain positive values of frequencies except the transverse acoustic mode near the $\Gamma$ point. This mode gets negative frequencies due the softening of phonons and a similar situation has been reported in the literature for the buckled germanene\cite{Ciraci}, where a strong dependence of frequency of this mode on the computational parameters is also observed. On the other hand, the planar arsenene is not stable since it possesses a few modes with imaginary frequencies in a large region of the Brillouin zone, which corresponds to negative values in Figure 1(h).  From the detailed analysis of the phonon spectra, we infer that, among the 12 phonon modes of the puckered arsenene, half of them are Raman active and they are 97, 112,  215, 217, 247 and 253 cm$^{-1}$ with the $C_{2h}$ point group symmetry at the $\Gamma$ point. In the case of the buckled arsenene, all the three modes of optical branch are Raman active. They are 236 cm$^{-1}$ (doubly degenerate) and 305 cm$^{-1}$ with the $D_{3d}$ point group symmetry at the $\Gamma$ point.

\section*{Band Structures.}
\quad

We show the electronic band structures for the puckered, planar and buckled arsenene in Figure 1(j),(k),(l). We find that the planar arsenene is metallic. We shall not continue to discuss any of its properties since it does not correspond to a stable structure. 
Both the puckered and buckled arsenene are semiconductors with indirect band gap of 0.831 and 1.635 eV, respectively. An indirect band gap in the buckled arsenene resembles that of the buckled (blue) phosphorene\cite{Blue}. However, there are certain differences in the band structures of the buckled phosphorene and arsenene. In the case of  the buckled arsenene (see Figure 1(i)), the valence band maximum lies at the $\Gamma$ point and the conduction band minimum occurs along the $\Gamma$-M direction, whereas in the buckled phosphorene, neither the conduction band minimum nor the valence band maximum lies at the high symmetry $k$-points in the Brillouin zone\cite{Blue}. Moreover, the difference between the indirect band gap ($1.635$eV) and the direct gap ($1.968$eV) at the $\Gamma$ point is quite large ($0.333$ eV) for buckled arsenene.

On the other hand,  the indirect-gap semiconducting character of the puckered arsenene is distinctly different from the direct band gap of the puckered phosphorene. We note that, in the puckered arsenene, two separate valence and conduction band edges exist near the Fermi energy. The competition between the energies of these edges crucially determines the nature of the semiconducting behavior. In the puckered arsenene, the maximum of valence and the minimum of conduction bands occur along the $\Gamma$-Y direction and  at the $\Gamma$ point, respectively. This causes the puckered arsenene to behave as an indirect-gap semiconductor. This is in contrast to the direct gap present at the $\Gamma$ point in the puckered phosphorene\cite{Peng}. Moreover, we clearly see from Figure 1(j) that the difference in energies between two valence band edges near the Fermi level is very small (only 85 meV) as compared to that in the puckered phosphorene (more than 500 meV)\cite{Peng}.  The difference is of the order of thermal energy at room temperature. In the next section, we analyze the effect of mechanical strain on the electronic structures of the puckered arsenene, and show that it is possible to transform puckered arsenene to a direct-gap semiconductor by applying strain.

\section*{Strain induced band gap change.}
\quad
\subsection{Puckered Arsenene}
\quad

Applying mechanical strain to the sample is a powerful method to modulate the electronic properties of materials. There are several reports which suggest that the band structure of phosphorene can be modified by applying strain\cite{Liu,Rodin,Fei,Peng}. In the present case, we study the evolution of electronic properties of the puckered arsenene when it is subjected to mechanical strain, both tensile and compressive, along the two separate lattice vectors $\textbf{a}$ and $\textbf{b}$. The application of mechanical strain is simulated by freezing one of the lattice constants, which is different from the optimized value and then vary the other lattice constant as well as internal degrees of freedom of each atom during the geometric optimization. Thus, the effect of strain is translated into the difference ($\Delta$a or $\Delta$b) between the frozen and globally optimized lattice constant.  For this purpose, we choose the range of $\Delta$a and $\Delta$b from $-20 \%$ to $20 \%$ of the optimized lattice constants with the spacing of $2 \%$. We assign the positive and negative values for compressive and tensile strains, respectively. First we evaluate the total energy of the puckered structure when strain is applied and the results are shown in Figure 2(a). Puckered and buckled arsenene are energetically stable under very strong compressive and tensile strains. The total energy with strain along the $b$-axis is lower than that with strain along the $a$-axis. It is natural that we can easily compress or expand the puckered structure along the $b$-axis by way of changing the puckered angle $\theta_1$. On the other hand, the structure is planar along the $a$-axis, which makes it difficult to change the structure along the $a$-axis. In Figure 2(b), we present the optimized lattice constants $a$ and $b$ during the compressive and tensile strains. For strain along the $a$-axis, we first fix the lattice constant $a$ and then optimize the lattice constant $b$. Similarly, we perform the optimization of  the lattice constant $a$ with fixed value of the lattice constant $b$ for the strain along the $b$-axis.

We also carry out detailed analysis of geometric structure of all the strained puckered arsenene. The results of this analysis are plotted in Figure 2(c) and (d) for strains along the $a$ and $b$ axes, respectively. In the case of the puckered structure, there are two types of bond lengths ($d_1$ and $d_2$) and bond angles ($\theta_1$ and $\theta_2$). Around the optimized structures, both the bond lengths and the bond angles vary linearly with the amount of strains. The increase in one of the bond lengths (or angles) results in decrease in the other bond lengths (or angles). This is due to the fact that the compression in one of the lattice directions leads to relaxation of atoms in the other direction.

Now, we discuss the modulations in the band structures of the puckered arsenene by applying strain along lattice vectors $\mathbf{a}$ (in Figure 3(a)) and $\mathbf{b}$ (in Figure 3(b)). We first investigate the band structures with strain along the $a$-axis. We find from Figure 3(a) that there is an indirect-to-direct gap transition due to both compressive and tensile strain along the $a$-axis. However, the locations of the direct band gap in $k$-vector for these two strains are different. In the case of compressive strain, the location of the direct band gap occurs at the $\Gamma$ point whereas for tensile strain, the direct band gap lies along the  $\Gamma$-Y direction. It is remarkable that the system remains as a direct-gap semiconductor for a wide values of strains ranging from $-10 \%$ to $10 \%$ except for the vicinity of no strain. This is a significantly important result from application point of view since 
it can accommodate possible structural  deformations, which may arise during growth or device manufacturing, while retaining its direct gap semiconducting behavior.

Other important observations from Figure 3(a) are the gap closing and the emergence of a linear dispersion around the Fermi energy along the $\Gamma$-Y direction for the tensile strain of $6\%$. In this situation, the system possesses a strong anisotropy in the electronic band structure;  the Dirac-like dispersion along the $\Gamma$-Y directions and the Schr\"{o}dinger-like dispersion in other direction.  In Figure 2(e) and (f), we plot the variation of the band gap as a function of strain. We observe that the band gap of the puckered arsenene becomes smaller when we apply tensile strain along the $a$-axis. The gap nearly closes at $6\%$ strain and then the band gap increases. On the other hand, for the compressive strain along the $a$-axis, the band gap initially decreases and becomes metallic around $-12\%$ due to the significant overlap between the conduction and the valence bands.

Next we investigate the band structures with strain along the $b$-axis. We find from Figure 3(b) that the application of strain along the $b$-axis produces nearly similar effects as in the case of strain along the $a$-axis, but in the opposite direction. It is natural because if we apply tensile strain along the $b$-axis, the system is elongated along the $b$-axis and shortened along the $a$-axis. This kind of deformation also occurs when we apply compressive strain along the $a$-axis. Here also we observe an indirect-to-direct band gap transition due to strain. Furthermore, the band gap of the puckered arsenene becomes smaller and then the gap nearly closes at $-10\%$ due to compressive strain. We note that the band structure of the puckered arsenene with $-10 \%$ compressive strain along the $b$-axis is almost similar to that with $6 \%$ tensile strain along the $a$-axis.

We find a strong correlation between the emergence of a linear dispersion in the band structure and the geometric structure of the puckered arsenene. A closer look at Figure 2(c) and (d) reveals that when the bond angles between the nearest neighbour become nearly equal ($\theta_1  \approx  \theta_2$), the system possesses  a nearly linear dispersion in the electronic band structure.
This can be understood as follows. When the two angles become equal, each arsenic atom is surrounded by three nearest neighbours with same angle. It makes the local environment more symmetric in spite of having non-hexagonal or non-trigonal crystal symmetries, which results in the gap closing with a nearly linear dispersion.

Our DFT based calculations predict that it is possible to produce the following modifications in the puckered arsenene by applying mechanical strains: (i) indirect-to-direct band gap transition, (ii) semiconductor-semimetal transition and (iii) semiconducting-metallic transition. Furthermore, the band gap of the puckered arsenene can also be tuned over wide range. These results suggest that the puckered arsenene can be choosen as one of promising nanomaterials for several applications, including optoelectronic devices.\\

\subsection{Buckled Arsenene}
\quad

Similar to the puckered arsenene, we have also performed the studies on the effect of strain on the properties of the buckled arsenene. In this case, we apply compressive and tensile strains symmetrically along its primitive lattice vectors. Then, the amount of  strain is directly quantified in terms of change in the lattice constant ($\Delta a$). Around the equilibrium geometric structure, the total energy of the system shows a parabolic behaviour, while the bond length and angle show nearly linear variations.

The band structure by varying strain is shown in Figure 4. The buckled arsenene mostly remains as an indirect-gap semiconductor for both compressive and tensile strains. The variation in indirect band gap with strain is plotted in Figure 2(f). The band gap slowly decreases with increase in either compressive or tensile strains. Then, the system goes from semiconductor to metallic for the values of strains beyond $-10 \%$ and $12 \%$.

\section*{Electric field induced band gap change.}
\quad

Applying perpendicular electric field to the buckled honeycomb structure such as silicene is shown to be an effective way to directly control the band gap\cite{Falko,EzawaNJP,Kamal}. In the buckled honeycomb structure, there exists a separation between the A and B sublattices. Then the perpendicular electric field acts as the staggered potential for the honeycomb system. It is an interesting problem to investigate how the band structure of the buckled arsenene is modified under electric field. We have performed calculations with the strength of perpendicular electric field ranging from 0.0 to 6.0 V/nm. It is found that there is no change in the band gap of the buckled arsenene for an electric field strength up to of 4.2 V/nm. The resultant band structures for some of the field strengths from 4.2 V/nm to 5.8 V/nm are shown in Figure 5. As discussed earlier, without electric field, the buckled arsenene is an indirect semiconductor. There is an indirect-to-direct gap transition at 4.2 V/nm.  We show the band gap as a function of electric field in Figure 5(b). For $4.1$ V/nm $ < E < 5.8 $ V/nm, the band gap decreases linearly with the strength of electric field. Based on the linear fit, we find that the band gap closes at the critical electric field $E=5.856$ V/nm. Above the critical field, the conduction band starts to overlap with the valence band and makes the system metallic. We note that the electric field required for the indirect-to-direct transition is quite high, which makes experiments difficult.

\bigskip
\noindent{\large\bf DISCUSSIONS}

We find that the bond lengths of these two-dimensional structures are less than that in bulk arsenic. 
In order to compare the stability of arsenene, we have also carried out the calculation of the stability on bulk grey arsenic. The buckling height is 1.291 \AA{} and angle is 96.72$^\circ$ for bulk grey arsenic. These values are very close to those of buckled arsenene.
It is very important to note that the cohesive energy of the puckered and buckled structures are quite close to that of bulk arsenic (-2.986 eV/atom). See Table 1. Hence, the growth of these two stable structures of arsenene are energetically favourable.

Graphene is manufactured by exfoliating graphite. Recently phosphorene has also been manufactured by exfoliating black phosphorus. There are a few reports on the existence of layered buckled arsenene in nature, which is called grey arsenic\cite{Grey, As-expt}. Thus, it is possible to obtain the buckled arsenene by exfoliating grey arsenic as in the case of graphene and phosphorene. Since the cohesive energy of the puckered arsenene is very close to both buckled arsenene and  bulk grey arsenic, there is also a possibility of manufacturing the puckered arsenene experimentally. Our results will motivate experimentalists to grow  arsenene.

We predict that the puckered and buckled structures of arsenene are stable from both the energetics and structural rigidity point of view
based on the DFT calculations. These two structures are semiconductors with indirect band gaps. Interestingly, the puckered arsenene goes from an indirect-gap to direct-gap semiconductor due to structural deformation along any of its lattice vectors. Furthermore, the onset of this transition occurs at very small amount of lattice deformation of $1\%$. It is also possible to tune the band gaps of this system over wide range while keeping its direct-gap semiconducting behavior by applying  compressive and tensile strain. Another important observation is the presence of the Dirac-like dispersion along the $\Gamma$-Y direction when the system is subjected to either compressive or tensile strain along its lattice vectors. For larger compressive strain, the system is transformed from a semiconductor to a metal due to a strong overlap of orbitals corresponding to the valence and conduction bands. Experimentally, it is possible to induce a strain by the beam bending apparatus\cite{Conley}, STM tips for tensile strain\cite{Lee} and substrates, and hence the results predicted can be verified once arsenene is grown on substrate or exfoliated from its bulk  counterpart.
The indirect-to-direct band gap transition found in puckered arsenene may open up a possibility of using this two-dimensional system in several optoelectronic devices such as a light-emitting diode and solar cell.

\bigskip
\noindent{\large\bf METHODS}
\section*{Computational Details}
\quad

We use Quantum ESPRESSO package\cite{QE} for performing a fully self-consistent density functional theory (DFT)\cite{DFT} calculations by solving the standard Kohn-Sham (KS) equations.  For exchange-correlation (XC) potential,  the generalized gradient approximation given by Perdew-Burke-Ernzerhof\cite{PBE} has been utilized. We use Vanderbilt ultrasoft pseudopotential\cite{Van} for As atom that includes the scalar-relativistic effect\cite{QE-lib}. Kinetic energy cutoffs of 30 Ry and 120 Ry have been used for electronic wave functions and charge densities, respectively. We adopt Monkhorst-Pack scheme for $k$-point sampling of Brillouin zone integrations with 31 $\times$ 31 $\times$ 1 and 31 $\times$ 21 $\times$ 1 for the buckled/planar and puckered systems, respectively. The convergence criteria for energy in SCF cycles is chosen to be 10$^{-10}$ Ry. The geometric structures are optimized by minimizing the forces on individual atoms with the criterion that the total force on each atom is below  10$^{-3}$ Ry/Bohr. In order to mimic the two-dimensional system, we employ a super cell geometry with a vacuum of about 18 \AA{}  in the z-direction (direction perpendicular to the plane of arsenene) so that the interaction between two adjacent unit cells in the periodic arrangement is negligible. The geometric structures are drawn using XCrySDen software\cite{XCR}

%SHOW THE PHASE DIAGRAMS AND EXPLAIN

\newpage

\bibliographystyle{naturemag}

\section*{Acknowledgements}
\quad
C.K thanks Dr. G. S. Lodha and Dr. P. D. Gupta for support and encouragement. C.K also thank the Scientific Computing Group, RRCAT for their support. M. E. thanks the support by the Grants-in-Aid for Scientific Research from the Ministry of Education, Science, Sports and Culture No. 25400317. C.K and M. E. are very much grateful to Dr. Aparna Chakrabarti and Prof. N. Nagaosa for many helpful discussions on the subject.

\section*{Contributions}
\quad
C.K performed the electronic structure calculations.
Both authors co-analysed the numerical results and co-wrote the manuscript.

\section*{Competing Interests}
\quad
The authors declare that they have no competing financial interests.

\section*{Correspondence}
\quad
Correspondence and requests for materials should be addressed to C. Kamal (ckamal@rrcat.gov.in) or Motohiko Ezawa (ezawa@ap.t.u-tokyo.ac.jp).

\begin{figure*}[t!]
\begin{center}
\includegraphics[width=0.9\textwidth]{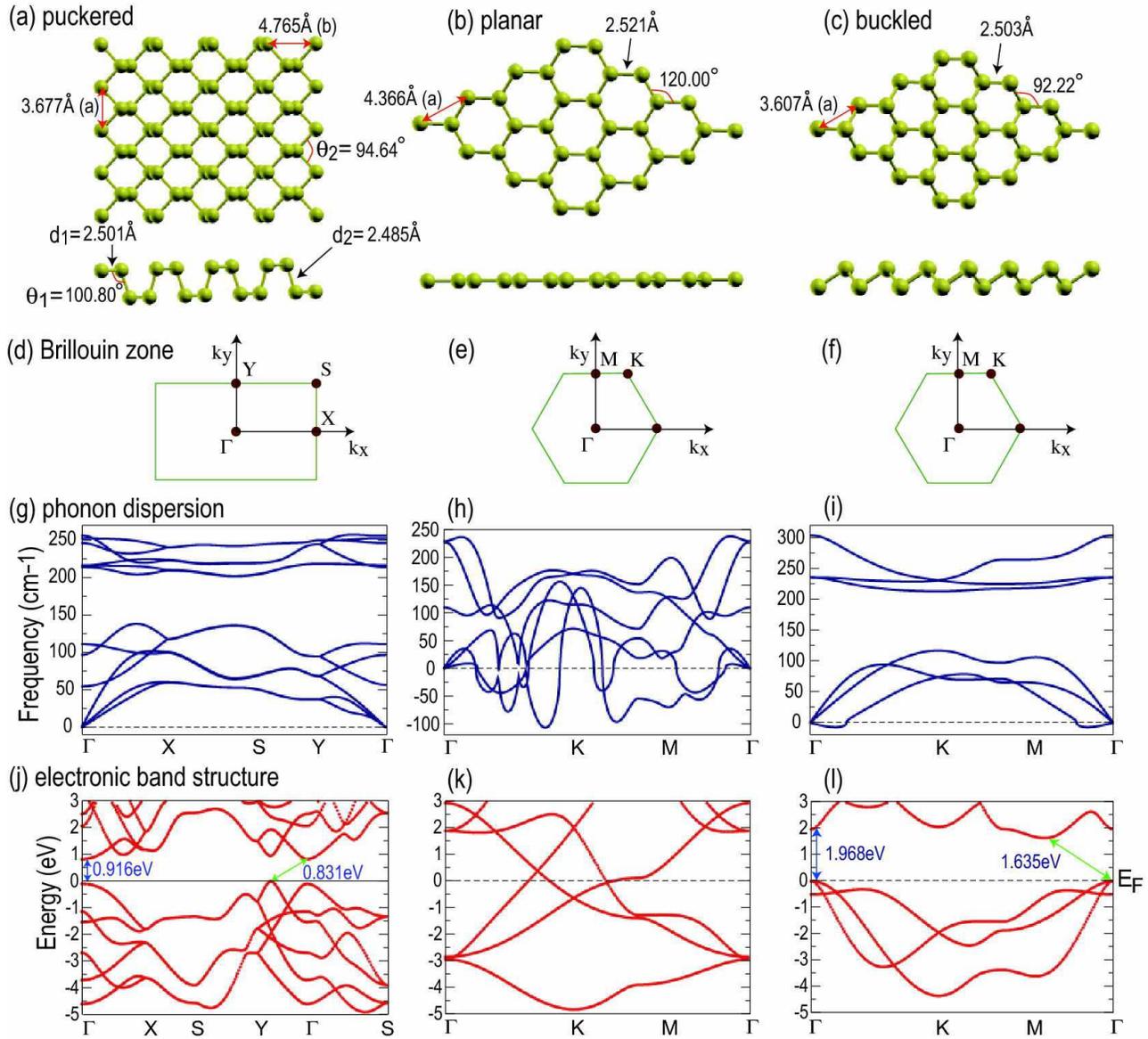}
\end{center}
\caption{\textbf{Optimized geometric structures, Phonon dispersion curves, Electronic band structure of puckered, planar and buckled arsenene.}
\textbf{(a),(b),(c)}: Fully optimized structure of \textbf{(a)} puckered, \textbf{(b)}  planar and \textbf{(c)} buckled arsenene. The length of arrow in red color indicates the lattice constant.
\textbf{(d)},\textbf{(e)},\textbf{(f)}: Brillouin zone of \textbf{(g)} puckered, \textbf{(h)} planar, and \textbf{(i)} buckled arsenene.
The Brillouin zone of puckered arsenene is rectangular, while those of planer and buckled arsenene are hexagonal. We mark the high symmetric points.
\textbf{(g)}, \textbf{(h)},\textbf{(i)}: Phonon dispersion curves for \textbf{(g)} puckered, \textbf{(h)} planar, and \textbf{(i)} buckled arsenene. The puckered arsenene is globally stable since the global minimum exists at the $\Gamma$ point. The planar arsenene is unstable since it possesses a few modes with negative frequency. In the buckled arsenene, all the modes contain positive values of frequency except transverse acoustic mode near the $\Gamma$ point.
\textbf{(j)},\textbf{(k)},\textbf{(l)}: Electronic band structure of \textbf{(j)} puckered, \textbf{(k)} planar, and \textbf{(l)} buckled arsenene. Indirect band gap (direct band gap at $\Gamma$) is indicated by green (blue) arrow. The puckered and buckled arsenene are indirect semiconductors, while the planar arsenene is a metal.
}
\end{figure*}

\begin{figure*}[t!]
\begin{center}
\includegraphics[width=0.9\textwidth]{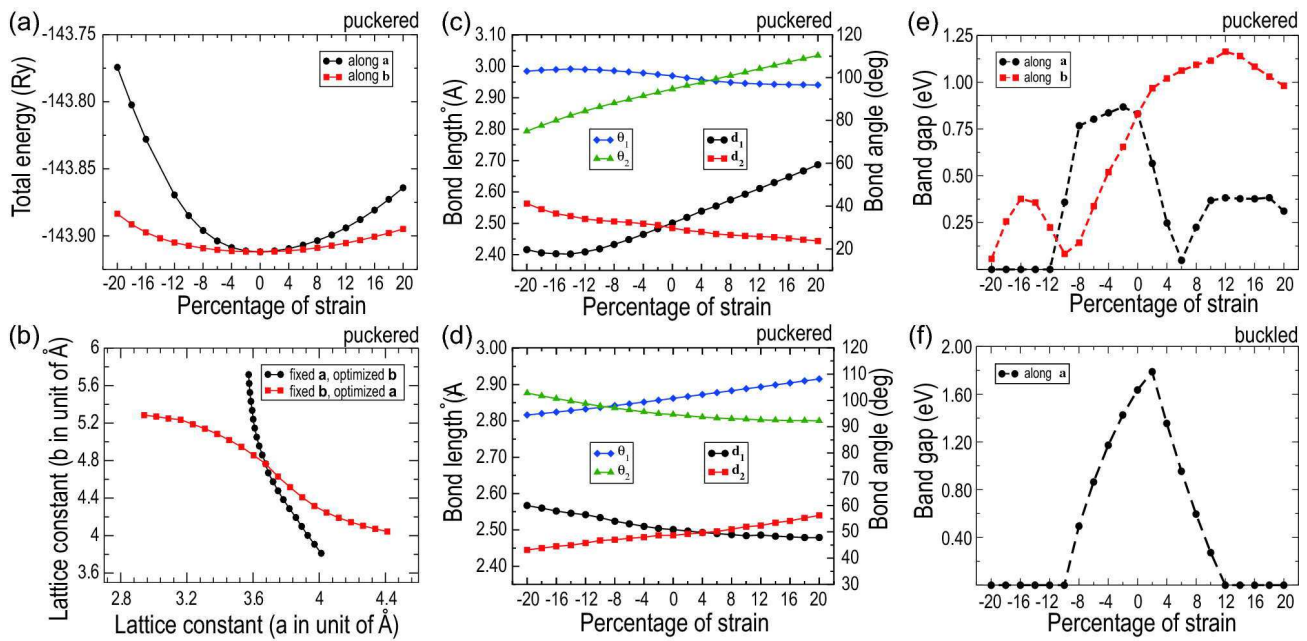}
\end{center}
\caption{\textbf{Effect of strain on energy, geometric structure and band structure.}
\textbf{(a)}: Variation of the total energy with strain along lattice vectors \textbf{a} and \textbf{b}  for puckered arsenene. The total energy is parabolic, where the bottom is at $0\%$ strain. The total energy with strain along lattice vectors \textbf{a} is higher than that along lattice vectors \textbf{b}.
\textbf{(b)}: Intersection of two curves represents the globally optimized lattice constants \textbf{a} and \textbf{b}. Black circles (red squares) show the optimized lattice constant \textbf{b} (\textbf{a}) by fixing the lattice constant \textbf{a} (\textbf{b}).
\textbf{(c),(d)}: Variation of bond lengths and bond angles with strain along lattice vectors \textbf{a} and \textbf{b}.
Around the equilibrium structure, the bond length and angle vary linearly with strain. The angles $\theta_1$ and $\theta_2$ become the same at $6\%$ ($-10\%$) strain along lattice vector \textbf{a} (\textbf{b}), where the band gap nearly closes. See Figure 3.
\textbf{(e)}: The band gap for puckered arsenene. The band gap reaches minimum at $6\%$ ($-10\%$) strain along lattice vector \textbf{a} (\textbf{b}), while it takes maximum at $-2\%$ ($12\%$) strain along lattice vector \textbf{a} (\textbf{b}). The band gap is zero beyond $-12\%$ strain along lattice vector \textbf{a}, where the system becomes metallic.
\textbf{(f)}: The band gap for buckled arsenene. The band gap attains  the maximum value at $2\%$ strain. The system becomes metallic beyond $12\% $  and $-10\% $ strain due to the overlap between the valence and conduction bands.
}
\end{figure*}

\begin{figure*}[t]
\begin{center}
\includegraphics[width=0.9\textwidth]{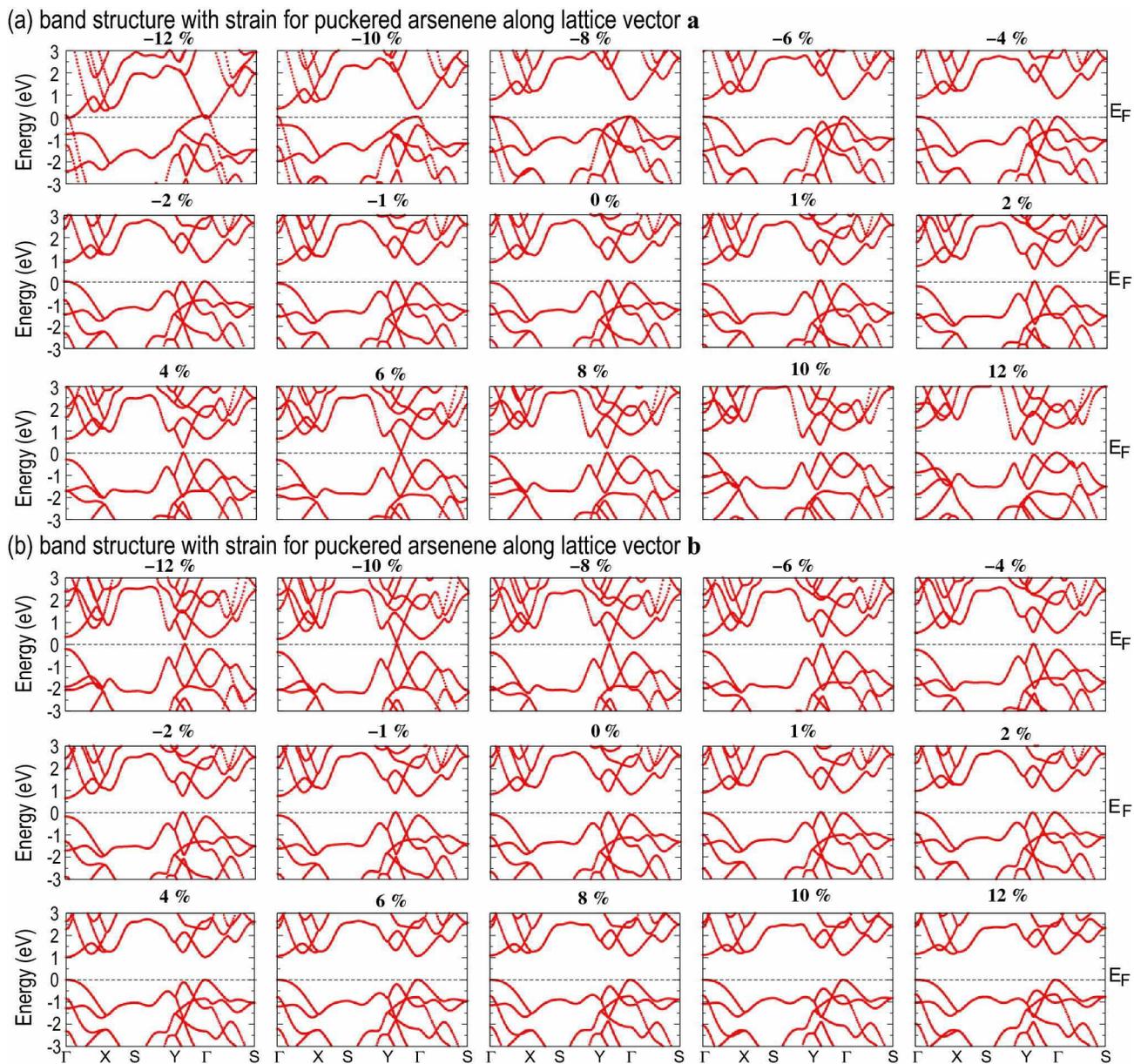}
\end{center}
\caption{\textbf{Variation of electronic band structures with strain along lattice vectors \textbf{a} and \textbf{b}.}
\textbf{(a)}: The band gap nearly closes at $6\%$ strain. The system is metallic beyond $-12\%$ strain. Indirect to direct gap transition occurs at $1\%$ and $-2\%$ strain.
\textbf{(b)}: The band gap nearly closes at $-10\%$ strain. Indirect to direct gap transition occurs at $2\%$ and $-4\%$ strain.}
\end{figure*}

\begin{figure*}[t]
\begin{center}
\includegraphics[width=0.9\textwidth]{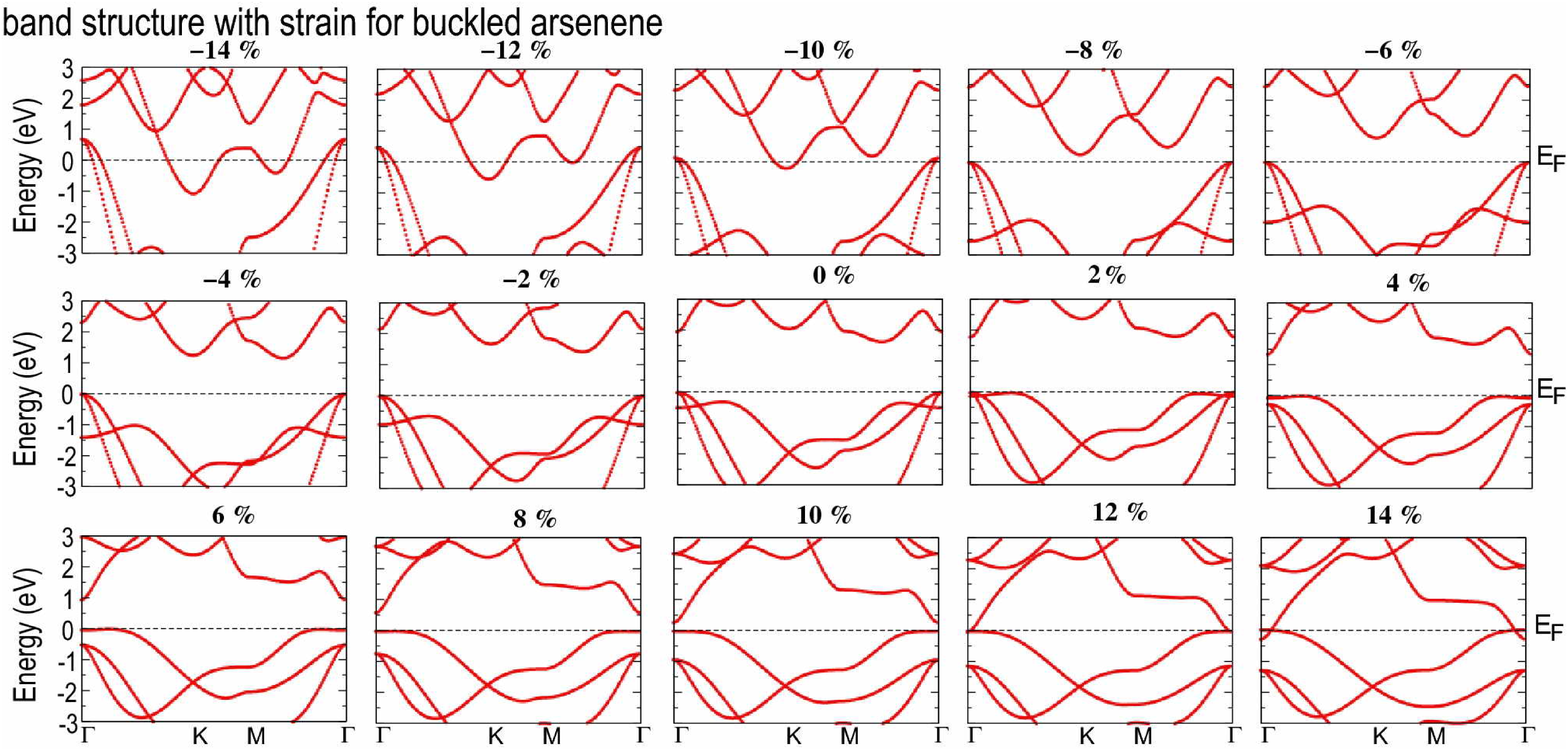}
\end{center}
\caption{\textbf{Variation of electronic band structures as function of strain for buckled arsenene.} The system becomes metallic beyond $12\%$ and $-10\%$ strain. 
}
\end{figure*}

\begin{figure*}[t]
\begin{center}
\includegraphics[width=0.9\textwidth]{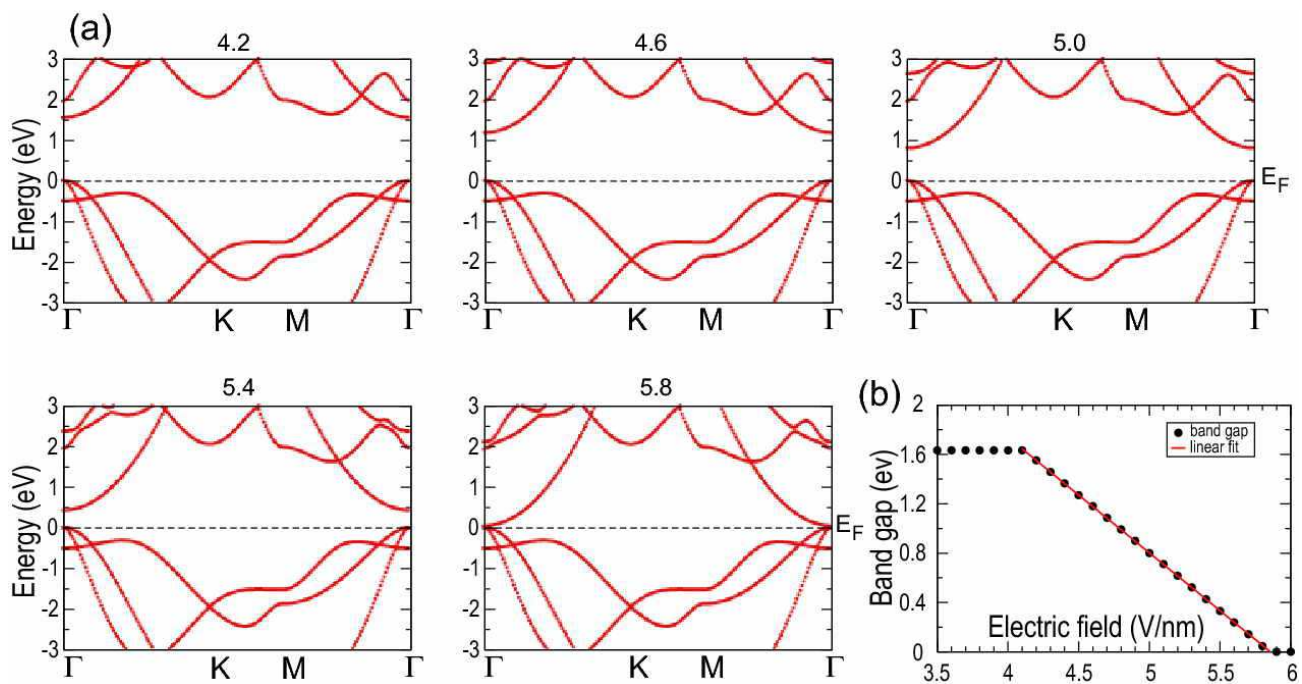}
\end{center}
\caption{\textbf{Variation of electronic band structures with the transverse static electric field for buckled arsenene.}
\textbf{(a)}: The value above the sub-figure represents the strength of the transverse electric field in unit of V/nm. The system becomes a direct-gap semiconductor beyond $4.2$V/nm and then becomes a metal beyond $5.8$V/nm due to the overlap between the valence and conduction bands. \textbf{(b)}: The evolution of the band gap with the strength of electric field. The band gap is constant below $4.2$V/nm, where the system is an indirect-gap semiconductor. The band gap linearly decreases between $4.2$V/nm and $5.8$V/nm. }
\end{figure*}

\end{document}